\documentstyle[aps,multicol,epsf]{revtex}

\begin{document}

\title{
Networks with desired correlations
}

\author{
S.N. Dorogovtsev$^{1, 2, \ast}$  
}

\address{
$^{1}$ Departamento de F\'\i sica and Centro de F\'\i sica do Porto, Faculdade 
de Ci\^encias, 
Universidade do Porto,\\
Rua do Campo Alegre 687, 4169-007 Porto, Portugal
\\
$^{2}$ A.F. Ioffe Physico-Technical Institute, 194021 St. Petersburg, 
Russia 
}

\maketitle

\begin{abstract} 
We discuss a simple method of constructing correlated random networks, which was recently proposed by M.~Bogu\~n\'a and R.~Pastor-Satorras 
(cond-mat/0306072). 
The result of this construction procedure is a sparse network whose degree--degree 
distribution asymptotically approaches a given function at large degrees. 
We argue that this convergence is possible if the desired function is sufficiently slowly decreasing.    
\end{abstract}


\begin{multicols}{2}

\narrowtext


Recently, Mari\'an Bogu\~n\'a and Romualdo Pastor-Satorras proposed a natural way of constructing correlated networks with desired correlations of the degrees of the nearest neighbors \cite{bp03}. 
This practical algorithm is based on the idea that previously was used in Refs.~\cite{gkk01,cl02,s02} for constructing uncorrelated graphs and in Ref.~\cite{ccrm02} for building correlated networks (see also Ref. \cite{pn03}). In all these constructions, (i) some weights (fitnesses, desired degrees, etc.) are ascribed to vertices, and (ii) each pair of vertices is connected with probability which depend on these weights.    

In the first version of this work, I reproduced this algorithm without knowing that it had already been proposed. So, in the present, 
corrected and shortened and more methodical version I mostly discuss the range of validity of the algorithm and features of networks, generated by this method.

The correlations between degrees of the nearest neighbors in a graph 
are naturally described by the joint distribution of the degrees of 
end vertices of an edge of the graph, $P(k,k')$,  
$P(k,k') = P(k',k)$,  
$\sum_{k,k'} P(k,k') = 1$. 
The joint degree--degree distribution determines the degree distribution 
$P(k)$ of a network: 
\begin{equation}
\sum_{k'} P(k,k') = \frac{kP(k)}{\overline{k}}
\, .   
\label{1}
\end{equation} 
Consequently, 
\begin{equation}
\overline{k} = \Biggl[ \sum_{k,k'} \frac{P(k,k')}{k} \Biggr]^{-1}
\,    
\label{2}
\end{equation} 
and $\langle k^n \rangle = \overline{k}\sum_{k,k'} k^{n-1} P(k,k')$. 

The algorithm of Bogu\~n\'a and Pastor-Satorras generates sparse random networks 
with desired degree--degree correlations. 
Suppose one wishes to obtain an ensemble of graphs with a desired joint 
distribution of the degrees of the nearest neighbors, $P(q,q')$. 
One should assume that $P(q,q')$ decreases with $q$ and $q'$ sufficiently 
slowly. 
Let the number $N$ of vertices in each graph of the ensemble be large and fixed. 

The procedure \cite{bp03} is as follows:  

\noindent
(i) Create $N$ vertices with a sequence of weights $\{q_i\}$, $i=1,\ldots,N$ independently sampled from the distribution 
\begin{equation}
P(q) = \overline{q}\sum_{q'} P(q,q')/q
\, ,    
\label{7}
\end{equation} 
where $\overline{q} = [\sum_{q,q'}P(q,q')/q]^{-1}$. 

\noindent
(ii) Put a link between $i$ and $j$ vertices with probability \cite{note1} 
\begin{equation}
p_{ij} = p(q_i,q_j) = \frac{\overline{q}}{N} \frac{P(q_i,q_j)}{P(q_i)P(q_j)}
\, .    
\label{8}
\end{equation} 

In this model, (i) the average degree of an $i$-th vertex coincides with $\overline{q}_i$, and (ii) the distribution of degrees of an individual vertex is a relatively narrow function at large degrees. 
This can be proved by calculating the degree distribution of a vertex with weight $q$ (see Ref.~\cite{bp03}). 
Here, alternatively, we simply find the first two moments, $\overline{k}_i$ and $\langle k^2_i \rangle$ for the degree of an $i$-th vertex.

The statistical weights of graphs $g \in G$ in the resulting 
ensemble, written in terms of $p_{ij}$ 
(adjacency matrix elements are $a_{ij} = 0,1$), are of a rather standard form  
\begin{equation}
\Pi(g) = \Pi(\{p_{ij}\},\{a_{ij}\}) \propto \prod_{i,j} \left (\frac{p_{ij}}{1-p_{ij}}\right)^{a_{ij}} \!= \prod_{i,j} s_{ij}^{a_{ij}} 
\,     
\label{9}
\end{equation} 
(compare with the statistical weights of classical random graphs and Ref.~\cite{pn03}). Here, $s_{ij} = p_{ij}/(1-p_{ij})$. 
The partition function of the ensemble is 
\begin{eqnarray}
Z(G)  =  Z(\{p_{ij}\}) & & = \sum_{g \in G} \Pi(g) = \sum_{a_{ij}} \prod_{i,j} \Pi(\{p_{ij}\},\{a_{ij}\}) \propto
\nonumber
\\[5pt]
& & \prod_{i,j} (1 - p_{ij})^{-1} = 
\prod_{i,j} (1 + s_{ij}) 
\, .    
\label{10}
\end{eqnarray} 
So, 
\begin{eqnarray}
& & \langle a_{ij} \rangle = 
\frac{\partial \ln Z(\{s_{ij}\})}{\partial \ln s_{ij}} 
= p_{ij} 
\, ,
\nonumber
\\[5pt]
& & 
\langle a_{ij}a_{i'j'} \rangle - \langle a_{ij} \rangle \langle a_{i'j'} \rangle = 
\frac{\partial}{\partial \ln s_{i'j'}}\,\frac{\partial }{\partial \ln s_{ij}} \ln Z(\{s_{ij}\}) = 
\nonumber
\\[5pt]
& & \phantom{WWWWWWWWWW} 
(p_{ij} - p_{ij}^2)\delta_{ii'}\delta_{jj'}
\, ,     
\label{11}
\end{eqnarray} 
etc. 
[The averages are over the statistical ensemble: 
$\langle X(g) \rangle = Z(G)^{-1}\sum_{g \in G}\Pi(g)X(g)$.] 
Consequently, all the moments are equal: 
$\langle a_{ij}^n \rangle = p_{ij}$, $n \geq 1$. This, in fact, is  
clear, since    
$a_{ij}$ takes only two values $0$ and $1$ with the probabilities $1-p_{ij}$ and $p_{ij}$, respectively. Then, 
\begin{equation}
\overline{k}_i  = \biggl\langle \sum_j a_{ij} \biggr\rangle = \sum_j p_{ij} = 
N\sum_q P(q) p(q_i,q) = \overline{q}_i 
\label{12}
\end{equation} 
[Eq. (\ref{8}) was used], so 
the mean degree in the net 
$\overline{k} = \overline{q}$, and 
\begin{equation}
\langle k_i^2 \rangle = 
\biggl\langle \biggl(\sum_j a_{ij}\biggr)^{\!2}\, \biggr\rangle = 
\sum_j p_{ij} + \biggl(\sum_j p_{ij}\biggr)^2 - \sum_j p_{ij}^2
\, .     
\label{13}
\end{equation} 

Note that Eqs. (\ref{12}) and (\ref{13}) may be obtained by using more naive arguments. Suppose for brevity that a net is of three vertices, $0$, $1$, and $2$. 
Then 
$\overline{k}_0 = 
1\cdot[p_{01}(1\!-\!p_{02})+(1\!-\!p_{01})p_{02}] + 2 \cdot p_{01}p_{02} = 
p_{01} + p_{02}$ 
[compare with Eq. (\ref{12})] 
and 
$\langle k_0^2 \rangle = 
1^2\cdot[p_{01}(1-p_{02})+(1-p_{01})p_{02}] + 2^2 \cdot p_{01}p_{02} = 
p_{01} + p_{02} + 2p_{01}p_{02}$ 
[compare with Eq. (\ref{13})]. 

Using the expression (\ref{8}), one can see that in a sparse network, the last term on the right-hand side of Eq. (\ref{13}) is of the order of $N^{-1}$ and so is negligible. Consequently, in the thermodynamic limit ($N \to \infty$), 
\begin{equation}
\langle k_i^2 \rangle - \overline{k}_i^2 = \overline{k}_i = \overline{q}_i  
\, .     
\label{14}
\end{equation} 
That is, the distribution of degrees of an individual vertex $i$ is indeed relatively narrow, if $k_i$ is large, and peaked at $\overline{q_i}$. 
One should take into account the fact that the number of edges in the net, which connect vertices with weights $q$ and $q'$, is 
\begin{eqnarray}
L(q,q') & = & \frac{1}{2}N^2 P(q)P(q') \langle a(q,q') \rangle = 
\nonumber
\\[5pt]
& & \frac{1}{2}N\overline{q}P(q,q') = L P(q,q')
\, ,     
\label{15}
\end{eqnarray} 
where $L$ is the total number of edges [note the symmetry factor $1/2$ in the second term of Eq. (\ref{15})]. Then one can     
finally conclude that the resulting degree--degree distribution of the net is expressed in terms of a given 
function: 
\begin{equation}
P_{\mbox{\scriptsize result}}(k,k') = P(k+\delta(k),k'+\delta'(k')) 
\, ,      
\label{16}
\end{equation} 
where the deviations, 
$|\delta(k)| \lesssim \sqrt{k},\ |\delta'(k')| \lesssim \sqrt{k'}$, are relatively small at large degrees. 
The relation (\ref{16}) shows that at large degrees the algorithm of 
Bogu\~n\'a and Pastor-Satorras provides the degree--degree distribution 
$P_{\mbox{\scriptsize result}}(k,k') \cong P(k,k')$, only if $P(k,k')$ decreases sufficiently slowly. The following arguments explain what this does mean. 

We ask two (related) questions. 
(i) How should a function $f(x)$ behave at large $x$ to satisfy 
the condition:  $f(x + c\sqrt{x})/f(x) \to 1$ as $x \to \infty$? 
(ii) How should a function $f(x)$ behave at large $x$ to guarantee that  
$[f(x + c\sqrt{x}) + f(x - c\sqrt{x})]/[2f(x)] \to 1$ as $x \to \infty$? Here, $c = \mbox{const}$. 

At first sight, the second, more symmetric condition may demand less strong restriction on the behavior of $f(x)$. 
However, both the questions have the same answer. Indeed, the second and the third, and the higher terms in the series 
$f(x + c\sqrt{x})/f(x) = 1 + c\sqrt{x} [df(x)/dx]/f(x) + 
c^2 x [d^2f(x)/dx^2]/[2f(x)] + \ldots$ approach zero at large $x$ if the same condition is satisfied: 
$f(x)$ must decrease with $x$ slower than $e^{-\sqrt{x}}$.   
So, if the given function $P(k,k')$ decreases slower than, say,   
$e^{-\sqrt{k}-\sqrt{k'}}$, then $P_{\mbox{\scriptsize result}}(k,k')$ asymptotically approaches $P(k,k')$ at large $k$ and $k'$ \cite{note2}. 
 

Moreover, this condition also guarantees that the resulting degree distribution $P_{\mbox{\scriptsize result}}(k)$ asymptotically approaches $P(k)$ at large degrees.  
On the other hand, in the region of small degrees, $P_{\mbox{\scriptsize result}}(k,k')$ deviates from desired $P(k,k')$, and $P_{\mbox{\scriptsize result}}(k)$ deviates from $P(k)$.  


Similar arguments are also valid for other network constructions of this type 
\cite{gkk01,cl02,s02,ccrm02,pn03}: desired distributions must decrease sufficiently slowly. 
On the one hand, this implies a serious restriction on the range of degree (or degree--degree) distributions which can be reproduced in such a way. 
On the other hand, it is the slowly decreasing distributions that are most interesting. 
\\

\noindent
{\small $^{\ast}$      Electronic address: sdorogov@fc.up.pt} 
\\

\end{multicols} 


\begin{references}
\bibitem{note1} Note that some of the probabilities $p_{ij}$ may turn out to be greater than $1$ 
\protect\cite{cl02,ccrm02,pn03}. This occurs at large $q_i$ and $q_j$ if 
$P(q,q')$ is slowly decreasing. 
To avoid such situations, one has to cut off the function at 
sufficiently small $q$ and $q'$. Alternatively, one can demand in 
Eq.~(\protect\ref{8}) that the probability $p_{ij}=1$, if $p(q_i,q_j) \geq 1$, 
which also produces a cut-off. The latter approach was in fact used in 
Ref.~\protect\cite{gkk01}.
\bibitem{note2} Even if we ``symmetrize'' the right-hand side of Eq. (\ref{16}) with respect to the deviations $\delta(k)$ and $\delta'(k')$ or present it in an integral form, the restriction on the behavior of $P(k,k')$ will not change.
\end{references}
\end{document}